\documentclass[prb,showpacs,showkeys,amsmath,amssymb]{revtex4}
\def\ba{\begin{eqnarray}}
\def\ea{\end{eqnarray}}
\usepackage{graphicx}
\usepackage{dcolumn}

\begin{document}
\title{On Phase Transition and Vortex Stability in the Generalized XY Models}
\author{L.A.S. M\'ol}\email{lucasmol@tdnet.com.br, lucasmol@vicosa.ufv.br}\author{A.R. Pereira}
\email{apereira@ufv.br} \affiliation{Departamento de F\'{\i}sica, Universidade
Federal de Vi\c{c}osa, 36570-000, Vi\c{c}osa, Minas Gerais, Brazil.}
\author{Winder A. Moura-Melo}
\email{winder@cbpf.br,winder@fafeod.br}
\affiliation{Departamento de Ci\^encias B\'asicas, Faculdades
Federais Integradas de Diamantina, Rua da Gl\'oria 187, 39100-000,
Diamantina, Minas Gerais, Brazil.}

\begin{abstract} We study a recent generalization proposed for the XY
model in two and three dimensions. Using both, the continuum limit
and discrete lattice, we obtained the vortex configuration and
shown that out-of-plane vortex solutions are deeply jeopardized
whenever the parameter of generalization, $L$, is increased. The
critical temperature for such models is calculated using the self
consistent harmonic approximation. In both, two- and
three-dimensional cases, such a temperature decreases with raising
$L$. Our results are also compared with other approximated
methods available in the literature.  \end{abstract}


\maketitle

\section{Introduction and Motivation}
The Hamiltonian $H=-J \sum_{<ij>} (S^x_i\,S^x_j +S^y_i\,S^y_j)$,
where $J$ is a coupling constant, the summation is taken over the
nearest-neighbors sites in a square lattice and $S^\alpha_i$
($\alpha=x,y$) is a spin component at site $i$, describes two
important models in magnetism known as the two-dimensional (2D)
planar rotator model (PRM) and the XY model. The difference
between these two models lies in the number of spin components.
Thus, in the PRM framework we have only two spin components which
are subject to the constraint $S^2_x+S^2_y=S^2$. On the other
hand, XY model also displays the third component with the
condition $S^2_x+S^2_y+S^2_z=S^2$. Hence, whenever fixing the spin
$S$ of the system, we are led with one and two independent spin
degrees of freedom for these models, respectively. Thus we may
parametrize the PRM physical spin content by only one scalar
field, the azimuthal angle $\varphi_i$, say,
$(\vec{S}_i)_{PRM}=S(\cos\varphi_i,\,\sin\varphi_i)$, while in the
XY model, we have two scalar, the azimuthal and polar angles, say,
$(\vec{S}_i)_{XY}=S(\sin\theta_i\cos\varphi_i,\,\sin\theta_i\sin\varphi_i,\,\cos\theta_i)$.
As it is well-known, the 2D PRM and XY models support topological
excitations and, although no long range order is established at
any finite temperature, they are shown to exhibit phase transition
related to the unbinding of vortex-antivortex pairs. Indeed, at
low temperatures, the vortex excitations are bound into pairs
whose interaction grows logarithmically with the distance between
the vortex centers. However, whenever the so-called
Berezinskii-Kosterlitz-Thouless (BKT) transition temperature,
$T_{BKT}$, is reached, the bound pairs appear to dissociate.

In addition, the static critical behavior of such models, mainly
the PRM, in two and three dimensions, has been studied for a long
time, using several statistical and Monte Carlo (MC) simulation
methods. Among others, such studies have led to the consensus upon
the nature of the phase transitions and the values of the critical
temperature and exponents \cite{nho,landau}. On the other hand, an
interesting extension of the 2D PRM was introduced around two
decades ago by Domany, Schick, and Swenden \cite{DSS}, in which
the potential is given by $V(\chi)=2J(1-[\cos^2(\chi/2)]^{p^2})$,
where $\chi\equiv \varphi_j-\varphi_i$. Clearly, the usual PRM is
recovered as long as $p=1$. Nevertheless, as $p$ is raised, an
increasingly narrow potential well, with a width around $\pi/p$,
is observed in the model. Notice also that whenever we have $\chi=
(2n+1)\pi$ ($n$ integer), then the potential gets a constant
value, $V((2n+1)\pi)=2J$. In addition, ordinary  spin-wave
behavior at low temperatures is ensured by the fact that, near the
minimum we do have $V(\chi)\approx Jp^2\chi^2/2$. It was also
shown by MC simulation that, for very large $p$ a first-order,
rather than a continuous phase transition takes place
\cite{himberg}. In this sense, it seems that the vortex
excitations alone are sufficient to account for both continuous
and first-order transitions, though in qualitatively distinct ways
\cite{himberg}.

Recently, a generalization of XY model has been proposed by Romano
and Zagrebnov \cite{romano}, whose potential looks like
$-J(\sin\theta_i\,\sin\theta_j)^L \,\cos(\varphi_i -\varphi_j)\,,$
with $L\in {\bf N}$. In their paper \cite{romano} the authors have
established rigorous inequalities holding for every $L\ge 1$.
Using Mean Field (MF) and Two-Site Cluster (TSC) techniques they
have also estimated the critical temperature, in the 3D case, for
some values of $L$. In the 2D case, and for arbitrary $L$, they
suggested that the above potential produces orientational disorder
at all finite temperatures, and supports BKT transition. However,
the stability and behavior of the vortex-like solutions in these
models are still unclear. Therefore, it would be important to know
about the types of vortices present in the system and their roles
in such a phase transition. Thus, in this work we present a
detailed analysis of the stability of vortex solutions. This is
done by applying discrete lattice and usual continuous
approximation approaches. The main conclusion is that out-of-plane
solutions are deeply jeopardized as long as the anisotropic
parameter, $L$, is increased. Furthermore, concerning the problem
of phase transitions it would also be desirable to verify in more
details the validity of some approximated methods used to estimate
the values of the critical temperatures. For this purpose, we
apply here the Self Consistent Harmonic Approximation (SCHA) to
the generalized XY model in two and three dimensions. Our results
strongly supports the idea that as $L$ is turned up the critical
temperature of the 2D and 3D generalized XY models appears to
decrease. However, it is worth noticing that our results predict
lower critical temperatures than those obtained by both, MF and
TSC methods. Namely, for the case $L=1$, SCHA approach yields the
value which is the closest to that found by means of Monte Carlo
simulations.

The generalized XY model is defined like below:
\ba\label{HXYgenrede}
H^{Gen}_{XY}=-J\sum_{<ij>}\Big[1-(S^z_i)^2 -(S^z_j)^2+(S^z_i S^z_j)^2
\Big]^{L/2}\, (S^x_i\,S^x_j +S^y_i\,S^y_j)
 \nonumber \\ =-J\,\sum_{<ij>}(\sin\theta_i\,\sin\theta_j)^L
\,\cos(\varphi_i -\varphi_j)\,, \ea where the spin vector at site
$i$ is defined by $\vec{S}_i=(S^x_i,S^y_i,S^z_i)$ satisfying the
non-linear constraint $\vec{S}^2_i=1$, and $L\in {\bf N}$.
Clearly, as long as $L=1$ we recover the usual XY-model, while for
$L=0$ we get the PRM (with $S^z=0$). In addition, as long as $L$
increases the anisotropy of the generalized model is taken up.
This Hamiltonian is a function of the spin components of the form
$f(S_i^z,S_j^z)[S_i^xS_j^x+S_i^yS_j^y]$, where
$f(S_i^z,S_j^z)=-[1-(S^z_i)^2 -(S^z_j)^2+(S^z_i S^z_j)^2 ]^{L/2}$.

\section{The solutions and their stability}

In order to study possible vortex-like and spin-wave
configurations, as well as their stability, we need to take into
account the continuous version of Hamiltonian (\ref{HXYgenrede}).
It is not difficult but length to obtain that: \ba
\label{HXYgencont} H^{\rm Gen}_{XY}=\frac{J}{2}\int d^2\vec{x}
\left[(1-m^2)^L(\vec{\nabla}\phi)^2
+L^2m^2(1-m^2)^{L-2}(\vec{\nabla} m)^2 -\frac{4}{a^2}(1-m^2)^L
+\frac{4}{a^2}\right]\,, \ea \noindent where $m=\cos\theta$, while
the constant $+4/a^2$ has been introduced
in order to renormalize the energy.\\

The equations of motion for spin-field variables,
$m(x,y;t)=\cos[\theta(x,y;t)]$ and $\phi(x,y;t)$ read like
follows: \ba & & \label{meq} \frac{\partial m}{\partial
t}=J(1-m^2)^{L-1}\left[2mL(\vec{\nabla}
m)\cdot(\vec{\nabla}\phi)-(1-m^2)\nabla^2\phi\right]\,,\\
& & \label{fieq}\frac{\partial \phi}{\partial
t}=JLm(1-m^2)^{L-2}\left[\frac{L(\vec{\nabla}m)^2}{(1-m^2)}-
\frac{L(L-2)m^2(\vec{\nabla}m)^2}{(1-m^2)^2}+
\frac{Lm\nabla^2m}{(1-m^2)} +(\vec{\nabla}\phi)^2-\frac{4}{a^2}
\right]. \ea Now, we shall pay attention to the possible
topological excitations associated to the non-linear equations of
motion, as well as their stability. First of all, we notice that
genuine static planar vortex-like configurations, $m=\cos\theta=0$
and $\phi=\arctan(y/x)$,
 are supported by the generalized model,
for every $L$ ($L\ge0$, $L$ integer). This can be easily checked
in Eqs. (\ref{meq}) and (\ref{fieq}). Later, let us notice that
whenever $m_i=0$, then $\sin\theta_i=+1$, implying that $L$ is
immaterial, since the energy experiences no changes as $L$ is
varied. Nevertheless, let us suppose quasi-planar configurations,
say, $m_i\approx 0$ and $\sin\theta_i\lesssim 1$. Now, the
scenario is deeply changed: indeed, increasing $L$ yields to
smaller effects of the term $(1-m^2)^L$ (or $f(S^z_i,S^z_j)$) in
the Hamiltonian, in a such a way that, as $L$ becomes very large,
this term practically vanishes, leading to an appreciable
decreasing of the spin interactions, and an increasing of the
system energy. Therefore, we may conclude that, as long as the
parameter $L$ is taken up, the out-of-plane spin components of the
generalized XY-model appear to become more absent in the system.
In other words, such components would require a greater amount of
energy to show up. It indicates that out-of-plane vortex solutions
must be unstable. The fact that out-of-plane fluctuations
$\langle (S_i^z)^2 \rangle$ must decrease as $L$ increases will be
used to calculate the critical temperature by means of the SCHA.

Now, in order to analyze the stability of possible out-of-plane
vortex solutions in more detail, we use the technique which has
been applied by Wysin \cite{wysin} to determine the critical
anisotropy parameter, $\lambda_c$, for the anisotropic Heisenberg
model~\cite{wysin,zaspel2,zaspel1}, whenever out-of-plane vortex
solutions become unstable~\cite{wysin,zaspel2}. This technique is
based upon the discrete lattice, and consists in analizing the
energy of out-of-plane vortex solutions, along with non-zero
out-of-plane spin components appearing only in a finite number of
sites equidistant from the vortex center. In a first
approximation, we consider that only the first four nearest spins
of the vortex center have non-zero out of plane components, $m_1
\neq 0$, while all the other sites display $m=0$. In order to
improve this calculation, we can include more sets of equidistant
sites. Considering the first three sets of equidistant spins at
radius $r_1=1/\sqrt{2}$, $r_2=\sqrt{10}/2$ and $r_3=\sqrt{18}/2$,
with out-of-plane components $m_1$, $m_2$ and $m_3$ respectively,
we then find the effects due to the $40$ bond sites nearest to the
vortex center: \ba E_3=-16J \left [
\frac{(1-m_2)^{L/2}}{5}+\frac{4(1-m_2)^{L/2}}{\sqrt{5}}+
\frac{(1-m_1^2+m_1m_2-m_2^2)^{L/2}}{\sqrt{5}} \right. \nonumber \\
\left. +\sqrt{\frac{2}{3}} (1-m_3^2)^{L/2}
+\frac{(1-m_2^2+m_2m_3-m_3^2)^{L/2}}{\sqrt{5}} - \frac{5}{2}
\right ], \ea where we have subtracted out the ground state energy
$-40J$ (in-plane exchange).

Minimization of the energy with respect to the out-of-plane spin
components demands the vanishing of the following derivatives:
\begin{subequations}\label{derivadas} \ba
\frac{\partial E}{\partial
m_1}=-\frac{8JL}{\sqrt{5}}(m_2-2m_1)(1-m_1^2+m_1m_2-m_2^2)^{L/2-1},
\ea \ba\frac{\partial E}{\partial
m_2}=-\frac{8JL}{\sqrt{5}}(m_1-2m_2)(1-m_1^2+m_1m_2-m_2^2)^{L/2-1}+
\frac{16JL}{5}m_2(1-m_2)^{L/2-1} \nonumber
\\ +\frac{64JL}{\sqrt{65}}m_2(1-m_2)^{L/2-1}
-\frac{8JL}{\sqrt{5}}(m_3-2m_2)(1-m_2^2+m_2m_3-m_3^2)^{L/2-1},\ea
\ba \frac{\partial E}{\partial
m_3}=-\frac{8JL}{\sqrt{5}}(m_2-2m_3)(1-m_2^2+m_2m_3-m_3^2)^{L/2-1}
\nonumber
\\+16JL\sqrt{\frac{2}{3}}m_3(1-m_3)^{L/2-1}. \ea
\end{subequations}
Then, out-of-plane vortex solutions are stable whenever
Eqs.(\ref{derivadas}) identically vanish, while $m_i\neq 0$, for
every $i=1,2,3$. To solve these equations, we assume small
amplitudes for $m_i$. In this case, the linearized version of
these equations reads like below:
\begin{subequations}\label{derivadas2} \ba
\frac{16JL}{\sqrt{5}}m_1-\frac{8JL}{\sqrt{5}}m_2=0, \ea \ba
-\frac{8JL}{\sqrt{5}}m_1+ \left(\frac{16JL}{5} +
\frac{32JL}{\sqrt{5}} +
\frac{64JL}{\sqrt{65}}\right)m_2-\frac{8JL}{\sqrt{5}}m_3=0,\ea
\ba -\frac{8JL}{\sqrt{5}}m_2+16JL
\left(\frac{1}{\sqrt{5}}+\sqrt{\frac{2}{3}}\right)m_3=0. \ea
\end{subequations}

Of course, trivial solutions, $(m_i=0)$, are always possible
(in-plane vortex). Nevertheless, our interest lies in the
conditions which could provide non-trivial configurations to show
up. The latter ones are possible if and only if the determinant of
the matrix coefficients is zero. However, such a determinant takes
the following value: $d\approx 3331.52(JL)^3$, which implies that
non-trivial configurations are possible only for $L=0$ (PRM, which
of course, can not support out-of-plane solutions). This result
clearly states us that out-of-plane vortex solutions are unstable.
Furthermore, we have also performed a similar plain for
equidistant sets of spins from $r_1=1/\sqrt{2}$ to
$r_6=\sqrt{50}/2$ (six sets of spins), whose conclusion is
analogous to the above one. In Ref.~[\onlinecite{wysin}], Wysin
found an accurate value of $\lambda_c$ (critical anisotropy) for
the anisotropic Heisenberg model which is in good agreement with
Monte Carlo results, using only three sets of equidistant spins
(form $r_1=1/\sqrt{2}$ to $r_3=\sqrt{18}/2$). Thus, we claim that
as long as we have used six sets of equidistant spins, our results
can be generalized leading us to conclude that out-of-plane vortex
solutions will be unstable in this model, (\ref{HXYgenrede}), for
every $L$ value.

Analogous conclusion is also obtained if we study the plane-wave
(meson) behavior of the model. In order to see that, let us take
expressions (\ref{meq}-\ref{fieq}), and suppose the appearance of
small time-dependent deviations from the planar vortex-like
solution ($m=m_0=\cos\theta=0$ and $\phi=\phi_0=\arctan(y/x)$),
like below: \ba
& & \label{mdev} m\longrightarrow m=m_0+m_1(\vec{x};t)=m_1(\vec{x};t)\,,\\
& & \label{fidev} \phi\longrightarrow
\phi=\phi_0+\phi_1(\vec{x};t)= \arctan(y/x)+\phi_1(\vec{x};t)\,.
\ea Since $m_1$ and $\phi_1$ are small, we may neglect higher
orders terms proportional to them and their derivatives in the
equations of motion~\cite{pereira2}. In this case we would be
neglecting the spin wave interactions. In addition, taking
$m_1(\vec{x};t)=m_1(\vec{x})\,e^{i\omega t}$ and
$\phi_1(\vec{x};t)=\phi_1(\vec{x})\,e^{i\omega t}$, where $\omega$
is the frequency of such long wavelength spin-wave excitations, we
get the following equations for $m_1$ and $\phi_1$: \ba
\label{eqm1} \nabla^2 \phi_1(\vec{x})=-\frac{i\omega}{J}m_1(\vec{x})\,, \\
\label{eqfi1} m_1(\vec{x})=
-\frac{i\omega}{JL\left(\frac{4}{a^2}-\frac{1}{r^2}\right)}\phi_1(\vec{x})\,,
\ea which imply that: \ba \label{eqfi1final}\nabla^2
\phi_1(\vec{x})=
-\frac{\omega^2}{J^2L\left(\frac{4}{a^2}-\frac{1}{r^2}\right)}\phi_1(\vec{x})\,.
\ea Then, as $r\to\infty$, we expect that such excitations look
like plane-waves, say,
$m_1(\vec{x})=m^0_1\,e^{i\vec{k}\cdot\vec{x}}$ and
$\phi_1(\vec{x})=\phi^0_1\,e^{i\vec{k}\cdot\vec{x}}$, getting: \ba
\label{omegaquad}\omega^2(|\vec{k}|;L)=\frac{+4J^2}{a^2}|\vec{k}|^2\,
L\,, \ea which confirms our earlier conclusion. Hence, for $L>1$
out-of-plane fluctuations appear to spend more and more energy. As
$L$ is increases, the magnon density decreases as $e^{-\sqrt{L}}$,
while the planar-vortex density is not affected. It may imply in
important differences in the thermodynamics of these Generalized
XY systems. In fact as has been shown by Currie {\it et.
al.}~\cite{currie}, the phase shift interaction between spin waves
and solitons provide the sharing mechanism of energy and degrees
of freedom among the nonlinear excitations of the system and
therefore is important in the study of the statistical mechanics
of the model.

\section{Phase Transitions in Two- and Three-Dimensional Generalized XY models}

Romano and Zagrebnov in Ref.~[\onlinecite{romano}] have shown that
this model supports Berezinskii-Kosterlitz-Thouless (BKT) phase
transition in 2D, and ordering transitions in 3D. In this section,
we perform an estimative of the critical temperature, in 2D and
3D, by applying the Self Consistent Harmonic Approximation (SCHA),
which has been widely appeared in the literature~\cite{ariosa1,
spisak,fishman, menezes, costa, pereira}, in connection with
isotropic and anisotropic models of the same class of universality
of the XY model.

Such a technique works by replacing the original Hamiltonian of the model
by a harmonic one, along with the coupling constant, $J$, by an effective one,
$K$, which takes into account the nonlinearities of the interactions.
First, we shall apply this technique to the spatially anisotropic 3D
generalized XY model, defined by the following hamiltonian: \ba
\label{3d} H_{3D}=-J\sum_{\bf{r,a}}\Big[1-(S^z_{\bf{r}})^2
-(S^z_{\bf{r+a}})^2+(S^z_{\bf{r}} S^z_{\bf{r+a}})^2 \Big]^{L/2}\,
\cos(\varphi_{\bf{r}}-\varphi_{\bf{r+a}}) \nonumber \\
-J_z\sum_{\bf{r,c}}\Big[1-(S^z_{\bf{r}})^2
-(S^z_{\bf{r+c}})^2+(S^z_{\bf{r}} S^z_{\bf{r+c}})^2 \Big]^{L/2}\,
\cos(\varphi_{\bf{r}}-\varphi_{\bf{r+c}}), \ea where $\bf{r+a}$
and $\bf{r+c}$ represents the nearest-neighbors in the XY plane,
and in the $z$ direction respectively, and $J_z$ is the
inter-plane coupling constant. The harmonic hamiltonian related to
the hamiltonian (\ref{3d}) is given by: \ba \label{h0}
H_0=\frac{1}{2}\sum_{\bf{r}} \Big[ K \sum_{\bf{a}}
(\varphi_{\bf{r}}-\varphi_{\bf{r+a}})^2 + K_z \sum_{\bf{c}}
(\varphi_{\bf{r}}-\varphi_{\bf{r+c}})^2  \nonumber  \\
 +(4LJ+2LJ_z)(S_{\bf{r}}^z)^2 \Big], \ea where $K$ and $K_z$
are given by: \ba K=J \left \langle \big[1-(S^z_{\bf{r}})^2
-(S^z_{\bf{r+a}})^2+(S^z_{\bf{r}} S^z_{\bf{r+a}})^2 \big]^{L/2}
\cos(\varphi_{\bf{r}}-\varphi_{\bf{r+a}}) \right \rangle, \\
K_z=J_z \left \langle \big[1-(S^z_{\bf{r}})^2
-(S^z_{\bf{r+c}})^2+(S^z_{\bf{r}} S^z_{\bf{r+c}})^2 \big]^{L/2}
\cos(\varphi_{\bf{r}}-\varphi_{\bf{r+c}}) \right \rangle.\ea These
averages
 are approximated replacing $\langle\ldots\rangle$ by
$\langle\ldots\rangle_0$, i.e., they are calculated using $H_0$
instead $H$. In this case, $\varphi_{\bf{r}}$ and $S_{\bf{r}}^z$
are decoupled allowing us to write: \ba \label{dec} K\approx
J\langle (1-(S_{\bf{r}}^z)^2)^{L} \rangle
\langle \cos(\varphi_{\bf{r}}-\varphi_{\bf{r+a}})\rangle, \\
K_z\approx J_z\langle (1-(S_{\bf{r}}^z)^2)^{L} \rangle \langle
\cos(\varphi_{\bf{r}}-\varphi_{\bf{r+c}})\rangle.\ea As we have
seen in Section II, as $L$ increases, the out-of-plane spins
fluctuations decreases. Using this result, we expand
$(1-(S_i^z)^2)^{L}$ as $1-L(S_i^z)^2+\ldots$, keeping only the
terms of the order of $(S_i^z)^2$. Then, the first average is
easily solved giving~\cite{costa, menezes}: \ba \label{sz}
1-L\langle(S_i^z)^2 \rangle = 1-\frac{T}{4J+2J_z}L . \ea

Hamiltonian (\ref{h0}), except for the $L$ term in
$(S_{\bf{r}}^2)$, has the same form of the Hamiltonian of
Ref.~[\onlinecite{costa}]. So, for a generic value of $L$, we
obtain, in the two limits $J_z \approx J$ and $J_z\ll J$, the
following expressions:

(a) $J_z \approx J$ \ba K=J\left(1-\frac{T}{4J+2J_z}L \right)\exp
\left \{-\frac{T}{16} \left[ \frac{3K+K_z}{2K^2+KK_z}+
\frac{3K+K_z}{2K^2+2KK_z} \right. \right. \nonumber \\ + \left.
\left. \frac{1}{2K} \left( \frac{1}{1-g}- \frac{g\tanh
^{-1}(1-g)^{1/2}}{(1-g)^{3/2}} \right) \right] \right \},\ea \ba
K_z=J_z\left(1-\frac{T}{4J+2J_z}L \right)\exp \left
\{-\frac{T}{16} \left[ \frac{2K+2K_z}{K_z^2+2KK_z}+
\frac{2K+2K_z}{(K+K_z)^2} \right. \right. \nonumber \\ + \left.
\left. + \frac{1}{K} \left(  \frac{\tanh
^{-1}(1-g)^{1/2}}{(1-g)^{3/2}}-\frac{1}{1-g} \right) \right]
\right \},\ea where $g=K/K_z$.

(b) $J_z \ll J$ \ba K\approx J\left(1-\frac{T}{4J+2J_z}L
\right)\exp \left \{-\frac{T}{16K}
\left[\frac{5}{2}+\frac{1}{1+K_z/K}+\frac{1}{2+K_z/K}\right.
\right. \nonumber \\  \left. \left.+\left( \frac{\pi ^2
+12}{24\pi}\right) \left(\frac{K_z}{K}\right) \ln
\left(\frac{K_z}{K}\right) \right] \right \}, \ea \ba K_z \approx
J_z \left(1-\frac{T}{4J+2J_z}L
\right)\left(\frac{K_z}{K}\right)^{Td/16K}\exp \left
\{-\frac{T}{16K}
\left[\frac{2}{1+K_z/K}+\frac{1}{2+K_z/K}+c\right] \right \}, \ea
where $d=(\pi^2+12)/12\pi$ and
$c=(\pi/12)\ln(4/\pi)+(1/\pi)\ln\pi$.

These self-consistent equations for the coupled variables $K$ and
$K_z$ gives the effective coupling at each value of the
temperature. The critical temperature is reached whenever these
equations admit only the trivial solutions, say, $K=0$ and
$K_z=0$. In this case, numerical data obtained from these
equations can be approximately fitted by \ba T_c \approx \frac{4+2
\lambda}{e+L}, \ea where $\lambda=J_z/J$.

For $L=0$ and $L=1$ we recover previous results from Refs.
[\onlinecite{costa,pereira}], for the PRM and XY models,
respectively, which are known to be in good agreement with those
obtained via Monte Carlo. In table \ref{tabela}, results from
Monte Carlo simulations for the PRM~\cite{chui} and XY
models~\cite{costa}, and those obtained by Romano and
Zagrebnov~\cite{romano} via MF and TSC techniques are compared
with those calculated by SCHA, for the isotropic case. It is worth
noticing that our results generally yields critical temperatures
lower than those predicted by MF and TSC techniques. Moreover, for
the $L=1$ case (usual XY model) for which we have a more detailed
study, our result is the closest to that obtained by MC
simulation.

\begin{table}
\caption{\label{tabela}Mean-Field, Two-Site Cluster, Monte Carlo
and SCHA results for the critical temperature of the isotropic 3D
generalized XY model. }
\begin{ruledtabular}
\begin{tabular}{ccccc}
$L$&Monte Carlo&Mean Field&Two-Site Cluster&SCHA\\
\hline 0&$2.2\pm 0.05$&&&2.2073\\
1&$1.54\pm0.01$&2.0000&1.7130&1.6136\\
2&&1.6000&1.4389&1.2716\\
3&&1.3741&1.2827&1.0493\\
4&&1.2190&1.1779&0.8931\\
\end{tabular}
\end{ruledtabular}
\end{table}

\begin{figure*}
\includegraphics[height=13cm,keepaspectratio]{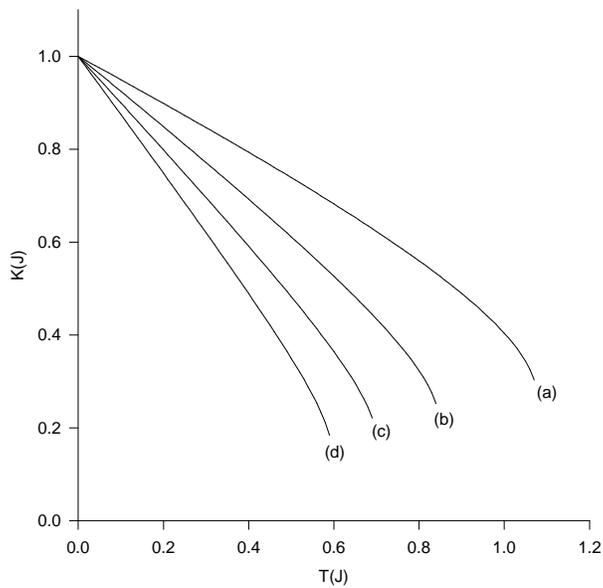}
\caption{\label{fig}Stiffness as a function of the temperature for
$L=1$ (a), $L=2$ (b), $L=3$ (c) and $L=4$ (d) in the two
dimensional case.}
\end{figure*}

On the other hand, whenever $J_z=0$, we get the 2D XY model. In
this case, the effective coupling constant, $K$, representing the
stiffness or superfluid density $\rho_s$, is given by \ba
K=J\left(1-\frac{T}{4J}L \right)\exp \left
(-\frac{T}{4K}\right).\ea Similarly, the critical temperature is
obtained whenever the self-consistent equation for $K$ is solved
only by taking $K=0$. Then, we found that the critical
temperature, as function of $L$, is $T=4/(e+L)$, which is quite
above Monte Carlo results for PRM and XY
models\cite{costa,pereira}. This occurs because the SCHA approach
attributes an excessive energetic cost to topological excitations,
such as vortex, causing an overestimation of the
Berezinskii-Kosterlitz-Thouless transition
temperature~\cite{costa,ariosa1}. In Fig.~\ref{fig} we show the
stiffness $K$ as a function of $T$ for some values of the
parameter $L$.

\section{Conclusions}In summary, we have obtained the continuum
limit for the Generalized XY-model, defined by Hamiltonian (1) and
studied the vortex stability. Calculations using the discrete
lattice confirmed the results obtained by the continuum approach.
Out-of-plane fluctuations, and as a consequence, the magnon
density, decrease as the parameter of generalization $L$
increases. The phase transition temperature as a function of $L$
was estimated using the self consistent harmonic approximation.
Our results seen to be more realistic than other approximated
methods available in the literature, such as two-site-cluster and
mean field. However, it has to be shown numerically and is beyond
the scope of this letter. We would like to consider in a future
paper some Monte Carlo simulations to calculate the critical
temperature for the Generalized XY-model in order to confirm our
assertion.

\begin{acknowledgments}
This work was partially supported by CNPq and Capes (Brazil).
\end{acknowledgments}

\end{document}